\documentclass[aps,pra,twocolumn,groupedaddress,superscriptaddress,amsmath,amssymb,,amsthm]{revtex4}
\usepackage{subfigure,hyperref,bbm,times}
\usepackage[T1]{fontenc}
\usepackage{braket}
\usepackage{amsthm}
\usepackage{epsfig}
\usepackage{color}
\usepackage{graphicx}% Include figure files
\usepackage{dcolumn}% Align table columns on decimal point
\usepackage{bm}% bold math

\newcommand{\scal}[2]{\langle#1|#2\rangle}
\providecommand{\openone}{\leavevmode\hbox{\small1\kern-3.8pt\normalsize1}}
\newtheorem{theor}{Theorem}

\begin{document}
\title{Universality of Schmidt decomposition and particle identity}

\author{Stefania Sciara} 
\affiliation{INRS-EMT, 1650 Boulevard Lionel-Boulet, Varennes, Québec J3X 1S2, Canada}
\affiliation{Dipartimento di Fisica e Chimica, Universit\`a di Palermo, via Archirafi 36, 90123 Palermo,
Italy}
\author{Rosario Lo Franco}
\email{rosario.lofranco@unipa.it}
\affiliation{Dipartimento di Fisica e Chimica, Universit\`a di Palermo, via Archirafi 36, 90123 Palermo,
Italy}
\affiliation{Dipartimento di Energia, Ingegneria dell'Informazione e Modelli Matematici, Universit\`{a} di Palermo, Viale delle Scienze, Ed. 9, 90128 Palermo, Italy}
\author{Giuseppe Compagno}
\affiliation{Dipartimento di Fisica e Chimica, Universit\`a di Palermo, via Archirafi 36, 90123 Palermo,
Italy}

\date{\today }% It is always \today, today,

\begin{abstract}
Schmidt decomposition is a widely employed tool of quantum theory which plays a key role for distinguishable particles in scenarios such as entanglement characterization, theory of measurement and state purification. Yet, it is held not to exist for identical particles, an open problem forbidding its application to analyze such many-body quantum systems. Here we prove, using a newly developed approach, that the Schmidt decomposition exists for identical particles and is thus universal. We find that it is affected by single-particle measurement localization and state overlap. We study paradigmatic two-particle systems where identical qubits and qutrits are located in the same place or in separated places. For the case of two qutrits in the same place, we show that their entanglement behavior, whose physical interpretation is given, differs from that obtained before by different methods. Our results are generalizable to multiparticle systems and open the way for further developments in quantum information theory when particle identity counts as a resource.
\end{abstract}

%\pacs{03.67.Mn, 03.65.Ud, 03.65.Ta}% PACS, the Physics and Astronomy

\maketitle

\section*{Introduction}

Systems of identical particles constitute the basic building blocks of quantum information theory, being present in Bose-Einstein condensates \cite{bloch2008many,anderlini2007controlled}, quantum dots \cite{kolodrubetz2009coherent,PhysRevLett.114.096602Tan,Martins2016PRL,Reed2016PRL}, superconducting circuits \cite{martinisfermions} and optical setups \cite{crespi2015,Morandotti2016Science}. Completely characterizing the quantum features of these composite systems is thus a crucial requirement from both fundamental and technological viewpoint.
Among these features, a tool which is at the heart of quantum information and quantum computation is provided by the Schmidt decomposition (SD) for bipartite systems of multilevel particles in pure states. It has general applications in entanglement characterization, theory of measurement, state purification, quantum erasure \cite{preskill1998lecture,nielsenchuang} and also in black-hole physics \cite{susskindbook,belokolos2009}. 
Despite its wide utilization in systems of distinguishable particles, the SD is claimed not to exist for identical particles \cite{tichy2011essential,balachandran2013entanglement}. Ordinarily, the SD unveils the entanglement of the system by the von Neumann entropy of the reduced density matrix, whose eigenvalues are the squares of the Schmidt coefficients appearing in the decomposition \cite{nielsenchuang}. It is nevertheless stated that ``the relationship between Schmidt coefficients and the eigenvalues of the reduced density matrix breaks down in the case of identical particles'' \cite{tichy2011essential}. 
As a consequence, the ordinary notion of partial trace to get the reduced state has been unsuited for assessing the entanglement in systems of identical particles \cite{tichy2011essential,balachandran2013entanglement,ghirardi2004general}, despite the importance of the latter as a quantum resource \cite{amico80entanglement,giovannetti2004Science,riedel2010Nature,benatti2014NJP,cramer2013NatComm,marzolino2015,eckert2002quantum,ghirardi2004general,horodecki2009quantum}. 

The problem of characterizing composite systems of identical particles stems from the usual first quantization formalism, where unobservable labels are assigned to particles to distinguish from one another \cite{cohen2005quantum,preskill1998lecture}. This practice induces fictitious correlations in the system due to the presence of these artificial labels, e.g. in the reduced density operator. 
There is a largely shared viewpoint that identical particle entanglement is just a formal artifact \cite{plenio2014PRL}. 
Identical particle entanglement has been a main issue and alternative methods have been developed to identify it  \cite{balachandran2013entanglement,plenio2014PRL,ghirardi2004general,schliemann2001quantum,eckert2002quantum,wiseman2003PRL,buscemi2007PRA,vogel2015PRA,giulianoEPJD,benatti2012AnnPhys,sasaki2011PRA,benatti2012PRA}. 

A particle-based approach was recently introduced that, without labelling identical particles, eliminates fictitious correlations \textit{ab ovo} and allows the use of ordinary notions based on the partial trace for identical particles \cite{lofranco2015quantum}.
Here we prove, within the aforementioned particle-based approach, the universality of the SD which also holds for pure states of two $d$-level identical particles, both bosons and fermions. This result extends all the applications of the SD known for distinguishable particles to identical ones, the Schmidt coefficients being always the square roots of the eigenvalues of the reduced density matrix. This SD gives the standard tool to straightforwardly characterize the entanglement for identical particles.

\section*{Results}

\textbf{Theory.}
We recall the notation of the intrinsically symmetric particle-based approach introduced in Ref.~\cite{lofranco2015quantum}. Hereafter, we mean by ``symmetric states'' (or ``symmetric Hilbert space'') the symmetric or antisymmetric behavior of the system states depending on the bosonic or fermionic nature of the particles, respectively. The overall state of two identical particles, one in the state $\phi$ and one in $\psi$, is completely characterized by enumerating the one-particle states and represented as $\ket{\phi,\psi}$. Two particles in $\ket{\phi,\psi}$ are not independent and their overall state is a whole which cannot be written as a tensorial product of one-particle states, i.e. $\ket{\phi,\psi}\neq \ket{\phi}\otimes\ket{\psi}$. However, a nonseparable external symmetric product of one-particle states (wedge product) can be introduced as $\ket{\phi,\psi}:=\ket{\phi}\times\ket{\psi}$. Analogously, we have $\bra{\phi,\psi}:=(\ket{\phi}\times\ket{\psi})^\dag=\bra{\psi}\times\bra{\phi}$ (this wedge product will be crucial in demonstrating the theorem below). The probability amplitude of finding the two particles in $\ket{\varphi,\zeta}$ if they are in $\ket{\phi,\psi}$, is obtained by the symmetric two-particle scalar product defined in terms of one-particle amplitudes as \cite{lofranco2015quantum}
\begin{equation}\label{amplitude}
\scal{\varphi,\zeta}{\phi,\psi}=\scal{\varphi}{\phi}\scal{\zeta}{\psi}+\eta\scal{\varphi}{\psi}\scal{\zeta}{\phi},
\end{equation}
where $\eta$ is $+1$ for bosons and $-1$ for fermions. This probability amplitude immediately shows that the generic state $\ket{\phi,\psi}$ is symmetric, i.e. $\ket{\phi,\psi}=\eta\ket{\psi,\phi}$. The state $\ket{\phi,\psi}$ spans a linear symmetric two-particle Hilbert space $H^{(2)}_\eta$. A symmetric inner product between state spaces of different dimensionality (one-particle projective measurement) can also be introduced as \cite{lofranco2015quantum} 
\begin{equation}\label{projection}
 \bra{\psi_k} \cdot \ket{\phi,\psi} \equiv \scal{\psi_k}{\phi,\psi}=\scal{\psi_k}{\phi}\ket{\psi}+\eta\scal{\psi_k}{\psi}\ket{\phi}.
\end{equation}
In $H^{(2)}_\eta$ it is possible to choose an orthonormal two-particle basis $ \{ \ket{i,j} \}$, $\ket{i}$ and $\ket{j}$ being single-particle states, where an arbitrary state of two identical particles can be expressed as $\ket{\Psi^{(2)}}=\sum_{ij}c_{ij}\ket{i,j}$. By Eq.~\eqref{projection}, one then gets the reduced (single-particle) density matrix via partial trace as \mbox{$\rho^{(1)}=\frac{1}{2}\sum_j\scal{j}{\Psi^{(2)}}\scal{\Psi^{(2)}}{j}=\frac{1}{2}\mathrm{Tr}^{(1)}\rho$} \cite{lofranco2015quantum}, where $\rho=\ket{\Psi^{(2)}}\bra{\Psi^{(2)}}$. We can now give the following theorem.

\begin{theor}
Within a symmetric two-particle Hilbert space $H^{(2)}_\eta$, a pure state of two $d$-level identical particles $\ket{\Psi}$ can always be written in the Schmidt decomposition (SD)
\begin{equation}\label{Schmidt decomposition}
\ket{\Psi}=\frac{1}{\sqrt{2}}\sum_i\sqrt{\lambda_i}\ket{i,\tilde{i}}.\quad {\displaystyle (\lambda_i > 0, \sum_i \lambda_i=1 )}
\end{equation}
The ``Schmidt coefficients'' $\sqrt{\lambda_i}$ are the square roots of the eigenvalues of the reduced density matrix and the states $\{ \ket{i} \}$ its eigenstates. The state $\ket{\tilde{i}}$ belongs to the basis $\{ \ket{i} \}$ and the symmetric two-particle basis  $\{ \ket{i,\tilde{i}} \}$ is the ``Schmidt basis''.
\end{theor}

\textit{Proof.} We express the state $\ket{\Psi}$ in terms of the symmetric two-particle basis $\{ \ket{i,j} \}$ as \mbox{$\ket{\Psi}=\frac{1}{2}\sum_{i,j}\ket{i,j}\scal{i,j}{\Psi}$}, where the symmetric two-particle identity matrix \mbox{$\mathbb{I}_2=\frac{1}{2}\sum_{i,j}\ket{i,j}\bra{i,j}$} has been inserted. By defining \mbox{$\ket{\bar{i}}\equiv\sum_j\scal{i,j}{\Psi}\ket{j}$}, the state can be further cast as $\ket{\Psi}=\frac{1}{2}\sum_i\ket{i,\bar{i}}$. Generally, the states $\{\ket{\bar{i}}\}$ are not orthonormal. Nevertheless, as for distinguishable particles \cite{preskill1998lecture}, there exists a basis $\{ \ket{i} \}$ where they are orthogonal, i.e. 
$\scal{\bar{i'}}{\bar{i}} \propto \delta_{ii'} $. We thus write
\begin{align} \label{matrix}
\scal{\bar{i'}}{\bar{i}}&=\sum_{j,j'}\bra{j'}\scal{\Psi}{i',j'}\scal{i,j}{\Psi}\ket{j} =\sum_{j}\scal{j,i}{\Psi}\scal{\Psi}{j,i'}\nonumber\\
&=\sum_j\bra{i}\times \scal{j}{\Psi}\scal{\Psi}{j}\times\ket{i'}=2\sum_{i,i'}\bra{i}\rho^{(1)}\ket{i'},
\end{align}
where we have used the partial trace $\rho^{(1)}=\frac{1}{2}\sum_j\scal{j}{\Psi}\scal{\Psi}{j}$. When the states $\{\ket{i}\}$ are the eigenstates of $\rho^{(1)}$, i.e. \mbox{$\rho^{(1)}\ket{i}=\lambda_i\ket{i}$}, the states $\{\ket{\bar{i}}\}$ are orthogonal and satisfy \mbox{$\scal{\bar{i}}{\bar{i'}}=2\lambda_i\delta_{ii'}$}. Denoting by $\{\ket{\tilde{i}}\}$ the set of orthonormal states associated to $\{\ket{\bar{i}}\}$, we have 
\begin{equation}\label{i tilde}
\ket{\tilde{i}}=\frac{1}{\sqrt{2\lambda_i}}\ket{\bar{i}}=\frac{1}{\sqrt{2}}\frac{1}{\sqrt{\lambda_i}}\sum_j\scal{i,j}{\Psi}\\ket{j}.
\end{equation}

Both $\ket{i}$ and $\ket{\tilde{i}}$ are eigenstates of $\rho^{(1)}$ with the same eigenvalue $\lambda_i$. Thus, given a set of eigenstates $\{\ket{i}\}$, each $\ket{\tilde{i}}$ is one of the states within the set. For bosons, if the eigenvalues are nondegenerate then $\ket{i}=\ket{\tilde{i}}$; for fermions, Pauli exclusion principle dictates $\scal{i}{\tilde{i}}=0$ and the eigenvalues are always degenerate (see Supplemental Material). Substituting $\ket{\bar{i}}=\sqrt{2\lambda_i}\ket{\tilde{i}}$ of Eq.~\eqref{i tilde} in $\ket{\Psi}=(1/2)\sum_i\ket{i,\bar{i}}$, the SD of Eq.~\eqref{Schmidt decomposition} is finally demonstrated. $\square$

When the states are characterized by more than one observable, for instance when the single-particle basis is $\ket{i}\equiv\ket{ab}=\ket{a}\otimes\ket{b}$ ($\ket{a}$ and $\ket{b}$ being two independent observables), one can be interested in studying the system for a fixed value of one of the observables. In such cases, the theorem above needs to be specialized. Let us take a two-particle state of the form $\ket{\Psi}=\ket{uv,u'v'}$, where $u$, $v$ are arbitrary single-particle states. 
This means that the SD of $\rho=\ket{\Psi}\bra{\Psi}$ is obtained by following the theorem above with the difference that the partial trace is now performed on the subspace of $b$ ($a$) with the observable $a$ ($b$) fixed. The corresponding reduced density matrix is indicated as $\rho^{(1)}_{a(b)}$ (see Supplemental Material). 
The universality of SD just proven entails its application to identical particles in many scenarios of quantum information (entanglement characterization, purification, measurement theory) in analogy with distinguishable particles \cite{nielsenchuang}. Knowledge of the Schmidt basis is essential to find the suitable set of measurements (Schmidt observables \cite{jaeger}) to acquire information on correlated identical particles in experimental contexts \cite{ekertAJP,Movassagh,PhysRevLett.103.110503}. 

\begin{figure*}[t]
\begin{center}
{\includegraphics[width= 0.94\textwidth]{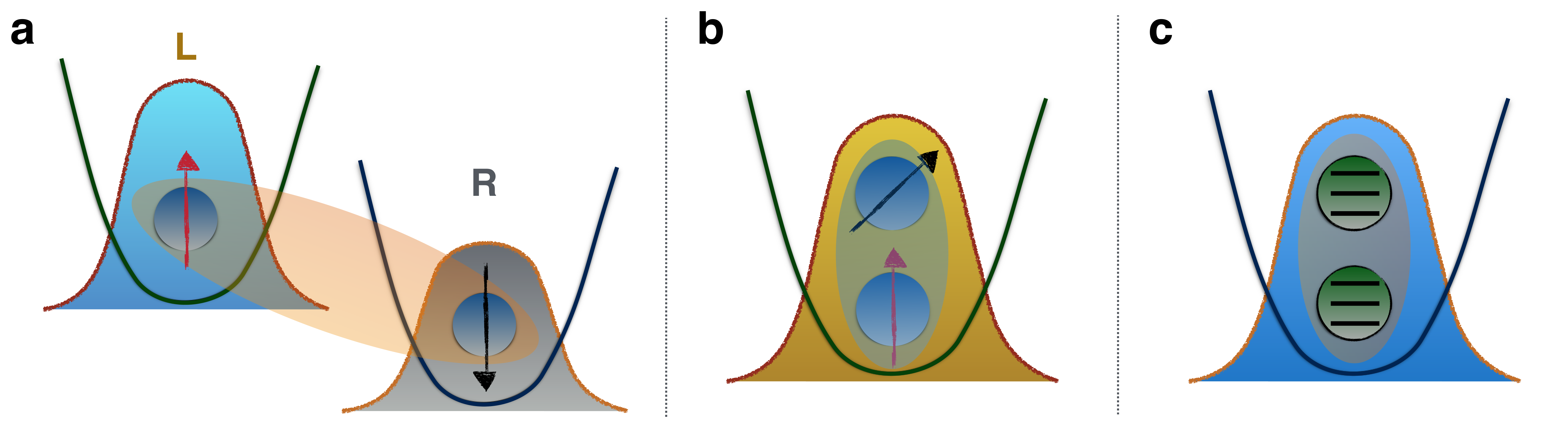}}
\end{center}
\caption{\small{\textbf{Illustrations of the studied systems.} (\textbf{a}) Two identical qubits in two spatially separated places with opposite pseudospins. (\textbf{b}) Two identical qubits in the same spatial mode with arbitrary pseudospins. (\textbf{c}) Two identical qutrits (three-level quantum systems) in the same spatial mode. The shaded ellipses indicate that the particles are entangled.}} 
\end{figure*}

The SD of Eq.~\eqref{Schmidt decomposition} defines an entangled state in terms of nonseparability, whatever the overlap between the particles. As for distinguishable particles \cite{preskill1998lecture}, we define the positive integer ``Schmidt number'' $s$ as the number of terms appearing in Eq.~\eqref{Schmidt decomposition}, that is the number of nonzero eigenvalues of $\rho^{(1)}$. If $s=1$, $\rho^{(1)}$ is pure and identifies a nonentangled state; if $s>1$, $\rho^{(1)}$ testifies an entangled state. The Schmidt number thus acts as entanglement witness. Analogous considerations hold for $\rho^{(1)}_{a(b)}$.
In particular, the (symmetric) basis state $\ket{i,j}$ with single-particle states $\ket{i}$, $\ket{j}$ containing only one observable results to be unentangled when $\ket{i}=\ket{j}$, while it is maximally entangled when $\scal{i}{j}=0$ (see Supplemental Material).
Being the Schmidt coefficients $\sqrt{\lambda_i}$ the square roots of the eigenvalues of the single-particle reduced state, they immediately lead to the von Neumann entropy
\begin{equation}\label{entropy}
S(\rho^{(1)})=-\mathrm{Tr}^{(1)}(\rho^{(1)}\log_2\rho^{(1)})=-\sum_i\lambda_i\log_2\lambda_i,
\end{equation}
as a quantifier of entanglement for identical particles, exactly as happens for nonidentical particles \cite{preskill1998lecture}.

Given any pure state $\rho=\ket{\Psi}\bra{\Psi}$, its SD is obtained in a recipe format as follows:

i. perform the trace of $\rho$ on a chosen single-particle basis to get the reduced single-particle density matrix $\rho^{(1)}$ (or $\rho^{(1)}_{a(b)}$);

ii. calculate eigenvalues, $\lambda_i$, and eigenstates, $\ket{i}$, of $\rho^{(1)}$ ($\rho^{(1)}_{a(b)}$);

iii.  construct the states $\ket{\tilde{i}}$ and express the state $\ket{\Psi}$ in terms of the Schmidt basis $\{ \ket{i,\tilde{i}} \}$.

\textbf{Applications.}
In the following, we apply this recipe to some states of interest (see Fig.~1). The first one is a situation already known \cite{lofranco2015quantum} which is here particularly useful to present how our method works. The other ones are new examples which evidence the usefulness of SD in finding novel entanglement features of identical particles.

\textbf{Two qubits in two separated sites (Bell-like state).} 
We consider two identical particles (bosons or fermions) with orthogonal internal degrees of freedom (pseudospins) located in separated sites, described by 
\begin{equation}\label{bell state}
\ket{\Psi}=\alpha\ket{L\uparrow,R\downarrow}+\beta\ket{L\downarrow,R\uparrow},
\end{equation}
where $\alpha^2+|\beta|^2=1$ ($\alpha$ real, $\beta=|\beta|e^{i\theta}$ with $\theta$ being the relative phase). The site $M$ (\textit{Left} ($L$) or \textit{Right} ($R$)) and the pseudospin $\sigma$ ($\uparrow$, $\downarrow$) are independent observables. The two sites are nonoverlapping, behaving thus as ``physical'' labels. The state of Eq.~\eqref{bell state} recalls Bell-like states \cite{tichy2011essential,lofrancoreview}. It permits us to discuss the role of local and nonlocal measurement in the structure of the SD for identical particle states. When local (single-particle) measurements of a particle property (e.g., the pseudospin) are performed in a localized region of space (e.g., $L$), the partial trace is \textit{local} \cite{lofranco2015quantum}. According to the recipe above, this measurement corresponds to project $\rho^{(1)}$ on the local basis, i.e. on the subspace $\{ \ket{L\uparrow}, \ket{L\downarrow} \}$. The reduced single-particle density matrix is
\begin{equation}
\rho^{(1)}_{L}=|\beta|^2\ket{R\uparrow}\bra{R\uparrow}+\alpha^2\ket{R\downarrow}\bra{R\downarrow}.
\end{equation}
It has eigenvalues $\lambda_1=|\beta|^2$, $\lambda_2=\alpha^2$, and eigenstates $\ket{1}=\ket{R\uparrow}$, $\ket{\tilde{1}}=\eta\ket{L\downarrow}$, $\ket{2}=\ket{R\downarrow}$, $\ket{\tilde{2}}=\eta\ket{L\uparrow}$, which define the Schmidt basis. We notice that $\scal{i}{\tilde{i}}=0$ and that the particle statistics is intrinsically included by the presence of $\eta$. The SD of $\ket{\Psi}$ is
\begin{equation}\label{SDlocal}
\ket{\Psi}=|\beta|\ket{1,\tilde{1}}+\alpha\ket{2,\tilde{2}},
\end{equation}
with von Neumann entropy
\begin{equation}
S(\rho^{(1)}_{L})=-\alpha^2\log_2\left(\alpha^2\right)-(1-\alpha^2)\log_2\left(1-\alpha^2\right).
\label{locale}
\end{equation}
This result coincides with the known von Neumann entropy for two distinguishable particles in a Bell-like state \cite{horodecki2009quantum}.

When nonlocal (one-particle) measurements are performed simultaneously on both sites ($L$ and $R$) where the particle has nonzero probability of being found, the trace is \textit{nonlocal}.  Operationally, it corresponds to perform the partial trace of $\rho$ on the global single-particle basis, i.e. on $\{ \ket{L\uparrow}, \ket{L\downarrow}, \ket{R\uparrow}, \ket{R\downarrow} \}$. Following the recipe above by a global partial trace, we get a SD of the Bell-like state $\ket{\Psi}$ different from Eq.~\eqref{SDlocal} (see Supplemental Material) leading to
\begin{equation}
S(\rho^{(1)})=-\alpha^2\log_2\left(\frac{\alpha^2}{2}\right)-(1-\alpha^2)\log_2\left(\frac{1-\alpha^2}{2}\right).
\label{globale}
\end{equation}

The difference between $S(\rho^{(1)}_L)$ and $S(\rho^{(1)})$ highlights the importance of measurement localization on the structure of the SD and in turn on the entanglement between two identical particles located in different sites.
To further clarify this aspect, we consider the particular case $\alpha=1$ when the state $\ket{\Psi}$ becomes $\ket{\Psi'}=\ket{L\uparrow, R\downarrow}$, which is unentangled \cite{lofranco2015quantum} since the particles are in separated sites and behave as uncorrelated distinguishable particles \cite{horodecki2009quantum,ghirardi2004general,amico80entanglement}. For this state, $S(\rho^{(1)})=1$ and $S(\rho^{(1)}_L)=0$. The result $S(\rho^{(1)})=1$ explains, in terms of nonlocality of the measurement, the same result obtained by using second quantization \cite{PhysRevA.64.042310,PhysRevA.64.054302}. For systems of identical particles, local single-particle measurements supply the intrinsic entanglement \cite{lofranco2015quantum}, whilst nonlocal measurements yield ``measurement-induced entanglement'' \cite{tichy2011essential,roch2014observation}. This feature must be contrasted with what happens for distinguishable particles, where single-particle measurements always address individual particles.

\begin{figure}[t]
\begin{center}
{\includegraphics[width= 0.45\textwidth]{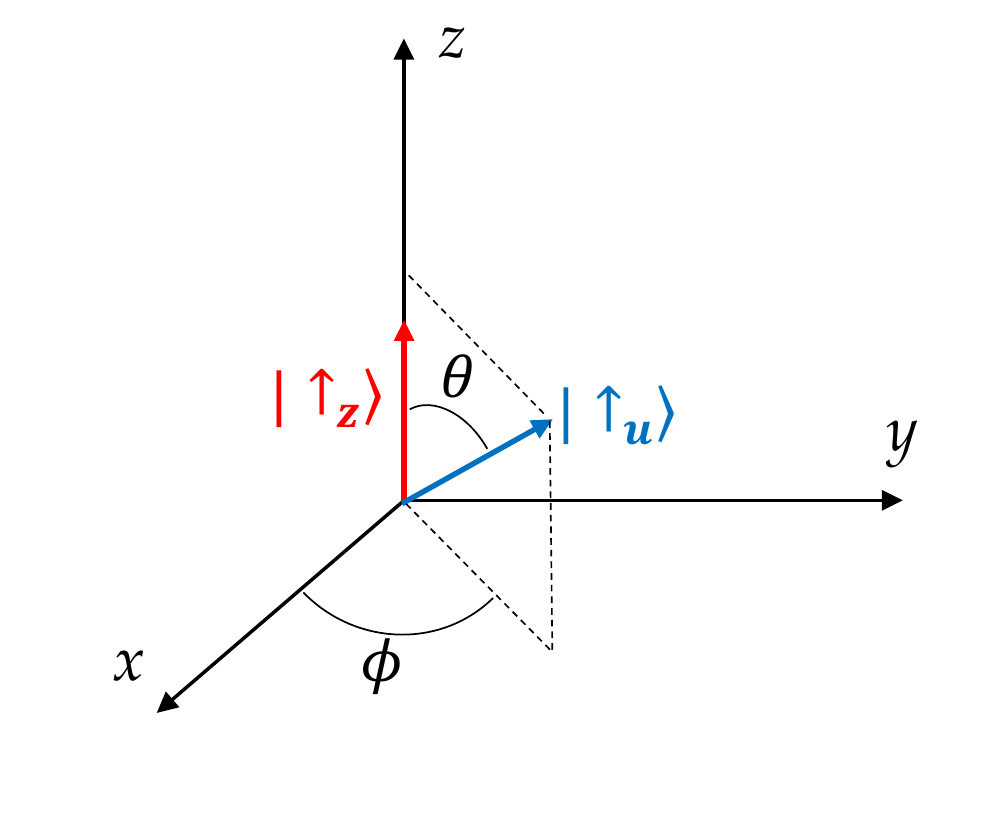}\vspace{-1cm}}
\end{center}
\caption{\small{\textbf{Geometric representation of the state of identical qubits in the same site with arbitrary pseudospins.} The two-qubit state is expressed by $\ket{\Phi}=\ket{\uparrow,\uparrow_\textbf{u}}=\cos\frac{\theta}{2}\ket{\uparrow,\uparrow}+e^{i\phi}\sin\frac{\theta}{2}\ket{\uparrow,\downarrow}$ ($\uparrow_z\equiv\uparrow$). One spin (red arrow) is along $z$-direction and the other (blue arrow) in the direction determined by the angles $\theta$ and $\phi$.}} 
\end{figure}

\textbf{Two qubits in the same site with arbitrary pseudospins.} 
Entanglement is a measure of nonseparability of the state \cite{modi2010unified}. When the particles are in the same site, their internal states (pseudospins) establish such nonseparability. A recent experiment showed that the entanglement in a Cooper pair can be extracted by means of graphene quantum dots, so that it can be possibly used as a resource for quantum information in the solid state \cite{PhysRevLett.114.096602Tan}. Moreover, it was recently observed \cite{kaufman2015entangling} that it is possible to prepare two maximally entangled ultra-cold atoms with opposing spin states by bringing them into the same optical tweezer (site). In general, one physically expects that situations may occur where particles are in the same site $M$ with pseudospins in arbitrary directions. Such a condition is possible only for bosons, since for fermions the only allowed state by the Pauli exclusion principle is that with opposite pseudospins which is maximally entangled \cite{lofranco2015quantum}. 
We hence study two identical boson qubits (e.g., photons) with one pseudospin along $z$-direction ($\uparrow_z\equiv\uparrow$) and the other one along the direction $\textbf{u}\equiv (1,\theta,\phi)$, as displayed in Fig.~2 (this situation generalizes that of two bosons with opposite pseudospins treated previously \cite{lofranco2015quantum}). By exploiting linearity and omitting the spatial index $M$, this state has the (unnormalized) form
\begin{equation}\label{generic spin}
\ket{\Phi}=\ket{\uparrow,\uparrow_\textbf{u}}=\cos(\theta/2)\ket{\uparrow,\uparrow}+e^{i\phi}\sin(\theta/2)\ket{\uparrow,\downarrow},
\end{equation}
where \mbox{$\ket{\uparrow_\textbf{u}}=\cos\frac{\theta}{2}\ket{\uparrow}+e^{i\phi}\sin\frac{\theta}{2}\ket{\downarrow}$}. Following the recipe above by performing the partial trace on the basis $\{\ket{\uparrow}, \ket{\downarrow} \}$ (see Supplemental Material), we obtain its (normalized) SD 
\begin{equation}
\ket{\Phi}=\sqrt{\frac{2}{\mathcal{N}}}\left(\cos^2\frac{\theta}{4}\ket{1,\tilde{1}}+\sin^2\frac{\theta}{4}\ket{2,\tilde{2}}\right),
\end{equation}
where $\mathcal{N}=1+\cos^2\frac{\theta}{2}$, $\ket{1}=\ket{\tilde{1}}=\cos\frac{\theta}{4}\ket{\uparrow}+\sin\frac{\theta}{4}\ket{\downarrow}$ and
$\ket{2}=\ket{\tilde{2}}=i(-\sin\frac{\theta}{4}\ket{\uparrow}+\cos\frac{\theta}{4}\ket{\downarrow})$. Notice the dependence of SD on $\theta$, which represents the pseudospin state overlap of the two particles. Entanglement of the two boson qubits is quantified by the von Neumann entropy $S(\rho^{(1)})=-(2/\mathcal{N})[\cos^4(\theta/4)\log_2(2\cos^4(\theta/4)/\mathcal{N})+\sin^4(\theta/4)\log_2(2\sin^4(\theta/4)/\mathcal{N})]$ and plotted in Fig.~3. It is maximum for $\theta=\pi$ (opposite pseudospins) and zero for $\theta=0$ (same pseudospins).

\begin{figure}[!t]
\begin{center}
{\includegraphics[width= 0.45\textwidth]{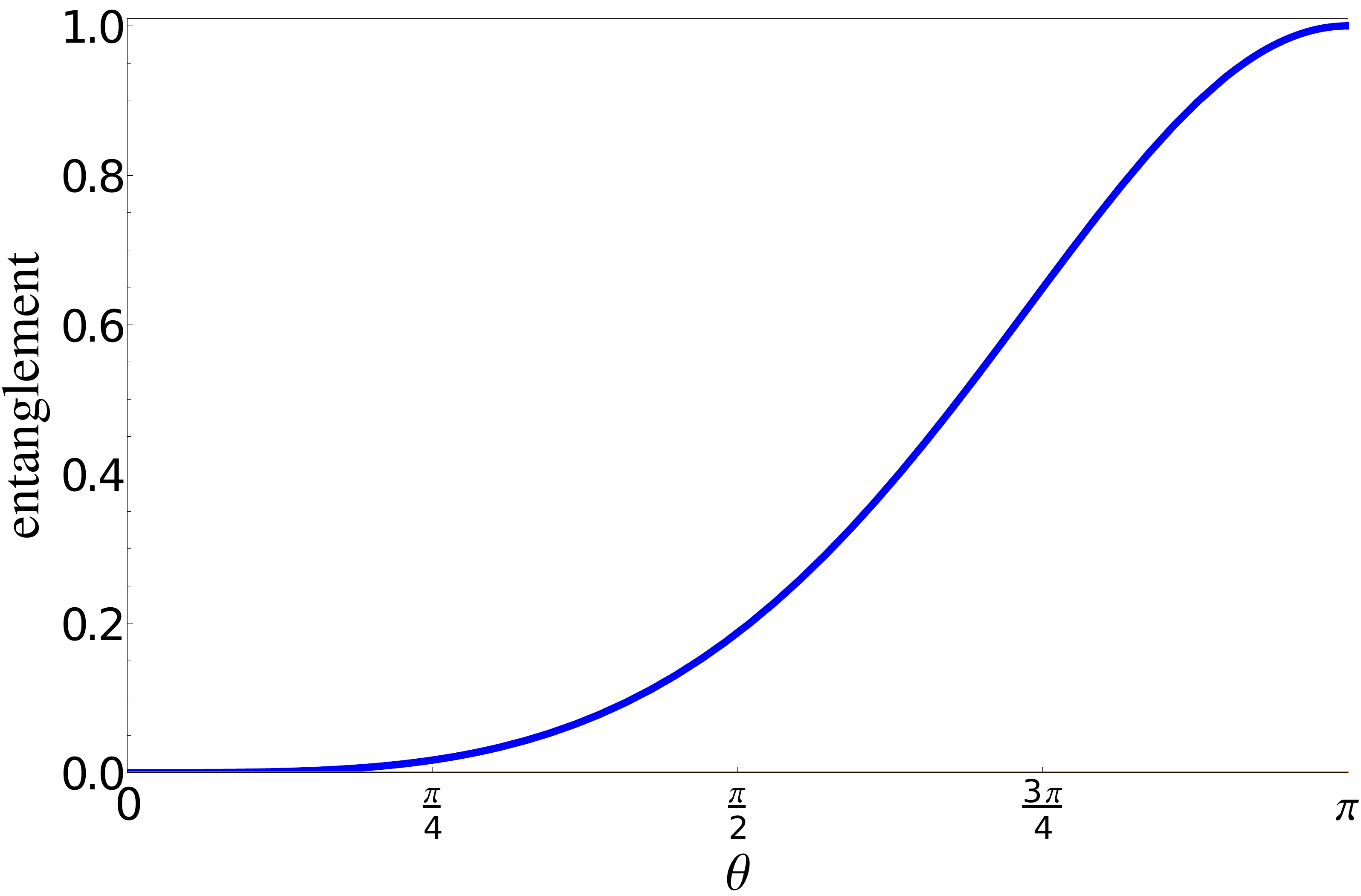}}
\end{center}
\caption{\small{\textbf{Entanglement behavior for identical bosons in the same site with arbitrary pseudospins.} Entanglement quantified by the von Neumann entropy of the state $\ket{\Phi}=\ket{\uparrow,\uparrow_\textbf{u}}$, \mbox{$\ket{\uparrow_\textbf{u}}=\cos\frac{\theta}{2}\ket{\uparrow}+e^{i\phi}\sin\frac{\theta}{2}\ket{\downarrow}$} is plotted as a function of $\theta$. }}
\end{figure}

\textbf{Two identical qutrits in the same site.} 
Systems of three-level particles (also called qutrits) are promising alternative candidates to be used in quantum processors instead of the standard two-level qubits \cite{PhysRevLett.100.060504,kumar}. Our approach applies to $d$-level identical particles (qudits) and allows us to analyze situations where this kind of systems are involved. We consider two identical qutrits in the same site, each characterized by the basis $ \{ \ket{e_1}, \ket{e_2}, \ket{e_3} \} $. This system is formally equivalent to that of two spin-1 bosons previously analyzed \cite{balachandran2013entanglement}. We take the state
\begin{equation}\label{stato balach}
\ket{\Psi_\phi}=\cos\phi\ket{e_1,e_2}+\sin\phi\ket{e_1,e_3},
\end{equation}
where the spatial index has been omitted for simplicity. 
By following the usual recipe (see Supplemental Material), we get its SD
\begin{equation}\label{schmidt balach}
\ket{\Psi_\phi}=\frac{1}{\sqrt{2}}\left(\frac{1}{\sqrt{2}}\ket{1,\tilde{1}}+\frac{1}{\sqrt{2}}\ket{2,\tilde{2}}\right),
\end{equation}
where $\ket{1}=\ket{\tilde{2}}=\cos\phi\ket{e_2}+\sin\phi\ket{e_3}$, $\ket{\tilde{1}}=\ket{2}=\ket{e_1}$, $ \ket{3}=-\sin\phi\ket{e_2}+\cos\phi\ket{e_3}$, $\ket{\tilde{3}}=0$. Expressing $\ket{\Psi_\phi}$ in the single-particle basis $\ket{i}$ ($i=1,2,3$) and exploiting the linearity of the symmetric Hilbert space \cite{lofranco2015quantum}, we get \mbox{$\ket{\Psi_\phi}=\ket{2,1}$}. 
The von Neumann entropy of Eq.~\eqref{entropy} is $S(\rho^{(1)})=1$, which represents a maximally entangled state independently of $\phi$. We provide a physical motivation to support this result. We notice that $\ket{\Psi_\phi}=\ket{2,1}=\ket{e_1,\phi}$, where \mbox{$\ket{\phi}=\cos\phi\ket{e_2}+\sin\phi\ket{e_3}$}. The independence of $\phi$ is due to the fact that the amount of entanglement only rests on the scalar product and hence on the angle between the single-particle states $\ket{e_1}$, $\ket{\phi}$, as depicted in Fig.~4. Moreover, entanglement is maximum because $\ket{e_1}$, $\ket{\phi}$ are orthogonal (as mentioned in \textit{Theory} section and demonstrated in the Supplemental Material). This situation is analogous to the case of two identical qubits in the same site with pseudospin states in arbitrary directions ($\ket{\uparrow,\uparrow_\textbf{u}}$) treated before. We observe that the state $\ket{\Psi_\phi}$ of Eq.~\eqref{stato balach} is obtained by specializing a state analyzed in the literature by subalgebra methods \cite{balachandran2013entanglement}. Our result ($\phi$-independent) contrasts with the previous one ($\phi$-dependent).

\begin{figure}[t!]
\begin{center}
{\includegraphics[width= 0.45\textwidth]{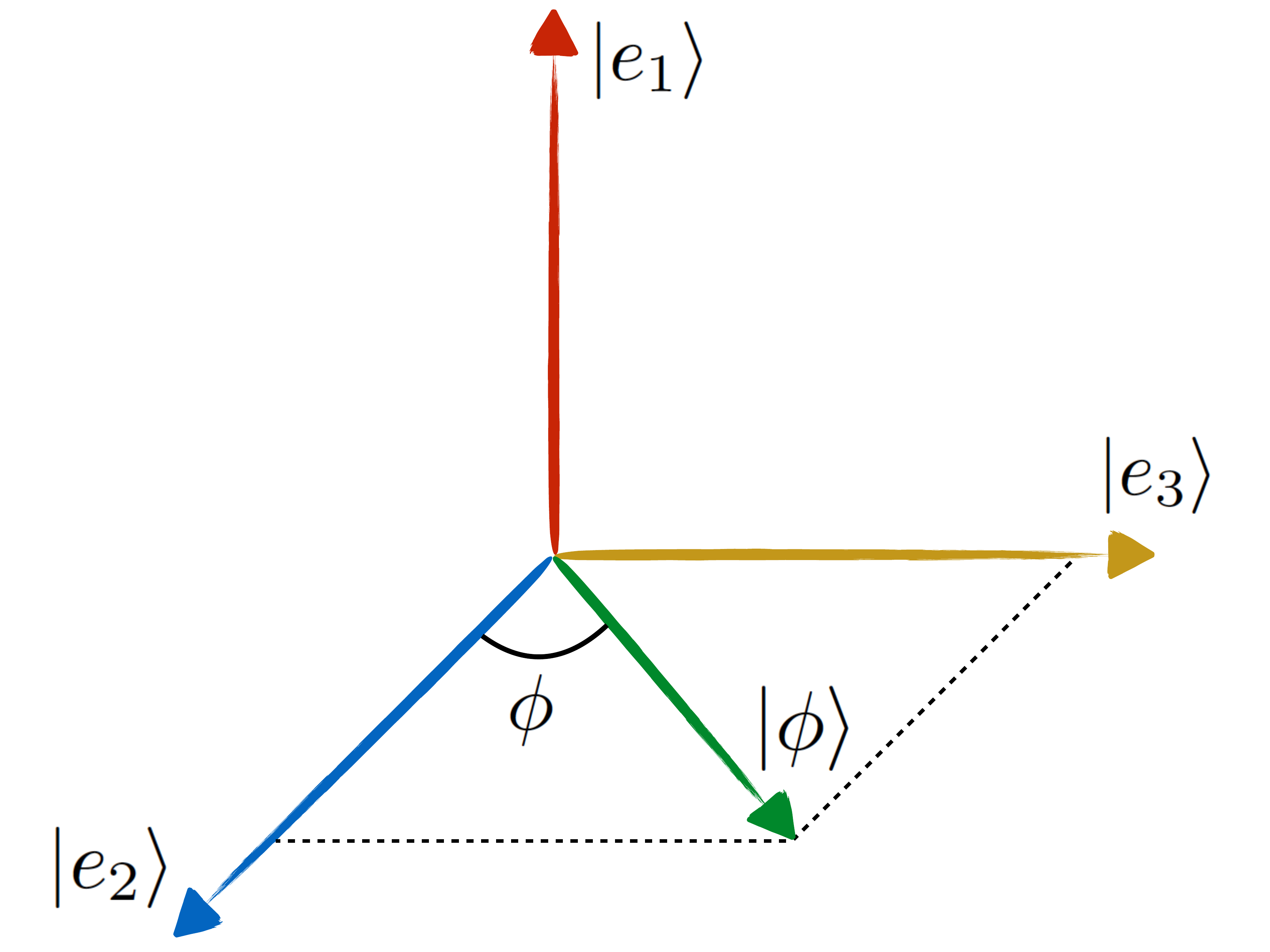}}
\end{center}
\caption{\small{\textbf{Geometric representation of the state of identical qutrits in the same site.} The two-qutrit state is expressed by $\ket{\Psi_\phi}=\cos\phi\ket{e_1,e_2}+\sin\phi\ket{e_1,e_3}=\ket{e_1,\phi}$, where $\ket{\phi}=\cos\phi\ket{e_2}+\sin\phi\ket{e_3}$. The single-particle states $\ket{e_1}$ and $\ket{\phi}$ are orthogonal.}}
\end{figure}

\section*{Discussion}

We have shown that, within a new approach in treating identical particles \cite{lofranco2015quantum}, the SD of bipartite quantum systems is universal, meaning that it can be defined for systems of both nonidentical and identical particles. This result goes against the common belief that the notion of SD does not exist for identical particles \cite{tichy2011essential} and constitutes a conceptual change in the field. We have found how the local and nonlocal nature of single-particle measurements, which define the partial trace operation, and the single-particle state overlap influence the structure of the SD. The Schmidt number maintains its role of entanglement witness while the Schmidt coefficients can be used to calculate the von Neumann entropy. 
These aspects also permit, differently from what has been claimed in the literature \cite{schliemann2001quantum,horodecki2009quantum,tichy2011essential}, to quantify entanglement of indistinguishable particles by ordinary notions. 

We observe that a difference exists from an operational point of view between nonidentical and identical particles. For distinguishable particles, SD and its corresponding entanglement are known to be exploitable within a resource theory by local operations, addressing each individual particle, and classical communication (LOCC) \cite{preskill1998lecture,nielsenchuang}. Differently, indistinguishable particles are not individually addressable. Nevertheless, the SD here proven for identical particles still allows its utilization by LOCC. In fact, this can be achieved by resorting to extraction procedures which make the overlapping identical particles tunnel with certain probabilities into two separated spatial modes \cite{plenio2014PRL}. For particles in the same site, where an intrinsic entanglement can be defined \cite{lofranco2015quantum}, it is straightforward to realize that the original Schmidt decomposition is reproduced, in a conditional fashion, into the two-particle state of the two accessible separated modes. These operational aspects will be treated elsewhere in major detail, including the case of partially overlapping identical particles, for which the definition of entanglement is more subtle \cite{lofranco2015quantum}.

We have applied the SD to analyze two boson qubits in the same site, finding how the amount of their entanglement depends on their pseudospin overlap: the entanglement increases as the two internal states tend to be orthogonal. This behavior generalizes previous results limited to orthogonal pseudospins \cite{lofranco2015quantum}. 
%which appear to confirm recent experimental observations of entanglement extraction in Cooper pairs \cite{PhysRevLett.114.096602Tan} and of entanglement generation between two cold atoms in the same optical tweezer \cite{kaufman2015entangling}
We have finally studied a system of two identical qutrits, which are relevant for storing quantum information \cite{PhysRevLett.100.060504,kumar}. We have straightforwardly obtained their entanglement and provided a physical interpretation. Our result differs from that determined for the same system by an alternative approach \cite{balachandran2013entanglement}. The origin of this difference in the entanglement measurement remains to be understood, requiring experimental verification and comparison of both theoretical approaches. 
 
Our result allows the natural generalization of the SD to arbitrary bipartitions of systems of $N$ identical particles. 
Our work enables the exploitation of this tool for characterizing composite quantum systems in theoretical and experimental relevant conditions where identical particles live in partially overlapping sites (e.g., electrons in quantum dots \cite{kolodrubetz2009coherent,PhysRevLett.114.096602Tan,Martins2016PRL,Reed2016PRL}, Bose-Einstein condensates \cite{bloch2008many}, solid-state qubits in circuit quantum electrodynamics \cite{martinisfermions} and wave-guided and integrated photons \cite{crespi2015,Morandotti2016Science}), which remain little explored. 
Our research demonstrates that entanglement between identical particles is not a mathematical artefact \cite{plenio2014PRL}, as has been argued, and provides methods to exploit the resources of entanglement comprised within identical particles for applications such as state teleportation, quantum metrology and quantum cryptography.

\textbf{Acknowledgements.} The authors would like to thank Reinhard F. Werner for a comment on the operational role of identical particle entanglement as a resource in occasion of a conference.

\appendix

\section{Eigenstates $\ket{\tilde{i}}$ of the reduced density matrix}
Here we demonstrate that the states $\ket{\tilde{i}}$ are eigenstates of $\rho^{(1)}$ with eigenvalues $\lambda_i$, analogously to the eigenstates $\ket{i}$. We start by using Eq.~(5) of the manuscript and reminding that \mbox{$\rho^{(1)}=\frac{1}{2}\sum_{i'}\braket{i'|\Psi}\braket{\Psi|i'}$} to have
\begin{align}
\rho^{(1)}\ket{\tilde{i}}&=\frac{1}{2}\sum_{i',j}\frac{1}{\sqrt{2\lambda_i}}\braket{i'|\Psi}\braket{\Psi|i'}\Braket{i,j|\Psi}\Ket{j}\nonumber\\
&=\frac{1}{2}\frac{1}{\sqrt{2\lambda_i}}\sum_{i',j}\braket{i'|\Psi}\bra{i}\times\braket{j|\Psi}\braket{\Psi|j}\times\ket{i'}.
\end{align}
Since $\sum_j\braket{j|\Psi}\braket{\Psi|j}=2\rho^{(1)}$ and $\bra{i}\rho^{(1)}\ket{i'}=\lambda_i\delta_{ii'}$, we obtain
\begin{equation}
\rho^{(1)}\ket{\tilde{i}}=\frac{1}{\sqrt{2}}\frac{1}{\sqrt{\lambda_i}}\lambda_i\braket{i|\Psi}.
\end{equation}
At this point, inserting the two-particle identity matrix \mbox{$\mathbb{I}_2=\frac{1}{2}\sum_{i',j'}\Ket{i',j'}\Bra{i',j'}$} between $\bra{i}$ and $\ket{\Psi}$ and using Eq.~(2) of the main text to get $\braket{i|i',j'}=\delta_{ii'}\ket{j'}+\eta\delta_{ij'}\ket{i'}$, we find
\begin{align}
\rho^{(1)}\ket{\tilde{i}}&=\frac{\sqrt{\lambda_i}}{\sqrt{2}}\frac{1}{2}\left[\sum_{j'}\braket{i,j'|\Psi}\ket{j'}+\eta\sum_{i'}\braket{i',i|\Psi}\ket{i'}\right]\nonumber\\
&=\frac{\sqrt{\lambda_i}}{\sqrt{2}}\frac{1}{2}[2\sqrt{2}\sqrt{\lambda_i}\ket{\tilde{i}}],
\end{align}
where the last equality is due to Eq.~(5) of the manuscript and to the symmetry property $\bra{i',i}=\eta\bra{i,i'}$ ($\eta^2=1$). Hence, we conclude that 
\begin{equation}
\rho^{(1)}\ket{\tilde{i}}=\lambda_i\ket{\tilde{i}},
\end{equation}
that is what we intended to demonstrate. Notice that the states $\ket{\tilde{i}}$ belong to the basis $\{\ket{i}\}$ of the eigenstates of the reduced density matrix.

\section{Relationship between the eigenstates $\ket{i}$ and $\ket{\tilde{i}}$}
According to Eq.~(5) of the manuscript, one has
\begin{equation}
\braket{i|\tilde{i}}=\bra{i}\sum_j\frac{1}{\sqrt{2\lambda_i}}\braket{i,j|\Psi}\ket{j}
=\frac{1}{\sqrt{2\lambda_i}}\braket{i,i|\Psi}.
\end{equation}
Expressing $\ket{\Psi}$ by the SD of Eq.~(3) of the main text, we obtain
\begin{equation}
\braket{i|\tilde{i}}=\frac{1}{\sqrt{2\lambda_i}}\bra{i,i}\sum_j\frac{\sqrt{\lambda_j}}{\sqrt{2}}\ket{j,\tilde{j}}
=\frac{1}{2}(1+\eta)\braket{i|\tilde{i}},
\end{equation}
where we have used $\braket{i,i|j,\tilde{j}}=(1+\eta)(\braket{i|j}\braket{i|\tilde{j}}+\eta\braket{i|\tilde{j}}\braket{i|j})$ (see Eq.~(1) of the manuscript) and $\braket{i|j}=\delta_{ij}$. 

From the previous equation, it is immediately seen that for fermions, as expected, it is always $\braket{i|\tilde{i}}=0$, since two of them cannot occupy the same state (Pauli exclusion principle). The orthogonality of the eigenstates $\ket{i}$ and $\ket{\tilde{i}}$ implies that the eigenvalues $\lambda_i$ of the reduced density matrix for a state of two fermions must be degenerate. 
For states of two bosons, instead, both cases of degenerate and non-degenerate eigenvalues can occur. In particular, if the eigenvalues $\lambda_i$ of the reduced density matrix are non-degenerate, it immediately follows $\braket{i|\tilde{i}}=1$: the eigenstates $\ket{i}$ and $\ket{\tilde{i}}$ coincide. 
We stress that these properties are always true when the eigenvalues of the reduced density matrix are calculated within the complete single-particle basis (including all possible outcomes of the observables which define a single-particle state) or in the specific case when the single-particle state is described by an observable alone, which are the conditions assumed in proving the theorem of the manuscript. Wider scenarios arise when the reduced density matrix is instead calculated by fixing a given value of an observable. 

%Therefore, for bosons the Schmidt decomposition of Eq.~\eqref{Schmidt decomposition}, in case of single states, is \begin{equation} \ket{\Psi_\mathrm{bosons}}=\frac{1}{\sqrt{2}}\sqrt{\lambda_i}\ket{i,i}. \end{equation}

\section{Partial trace on a given subspace of an observable}
Let us consider a single-particle state $\ket{i}\equiv\ket{ab}=\ket{a}\otimes\ket{b}$ and a two-particle state $\ket{\Phi}=\ket{u v, u' v'}$, where $a$, $b$, $u$, 
$v$, $u'$, $v'$ are arbitrary states corresponding to two independent observables $A$ and $B$ (e.g., the site and the spin of the particle). We show a general criterion to perform the partial trace of $\ket{\Phi}\bra{\Phi}$ on the subspace of an observable (e.g., $a$) by varying the other one (e.g., $b$). We first calculate the one-particle projective measurement (see Eq.~(2) of the manuscript)
\begin{align}
\braket{ab|\Phi}&=\braket{ab|uv,u'v'} \nonumber \\
&=\bra{a}\otimes ( \braket{b|v}\ket{u}\times\ket{u'v'}+\eta\braket{b|v'}\ket{uv}\times \ket{u'} ).
\end{align}
The action of $\bra{b}$ on the state $\ket{\Phi}$ can be thus defined as
 \begin{equation}
\braket{b|uv,u'v'}=\braket{b|v}\ket{u}\times\ket{u'v'}+\eta\braket{b|v'}\ket{uv}\times \ket{u'}.
\end{equation}
The reduced single-particle density matrix performed on the subspace $a$ of the observable $A$ (that is, obtained by fixing $a$ and summing on $b$) reads
\begin{eqnarray}
\rho^{(1)}_a&=&\braket{ab|\Phi}\braket{\Phi|ab}=\braket{ab|uv}\braket{u'v'|ab}\nonumber\\
&=&\bra{a}\otimes \sum_b\{ \braket{b|v}\ket{u}\bra{u}\times\ket{u'v'}\bra{u'v'}+\nonumber\\
&&+\braket{b|v'}\ket{u'}\bra{u'}\times \ket{uv}\bra{uv}\nonumber\\
&&+\eta ( \braket{b|v}\braket{b|v'}\ket{u}\bra{u'}\times \ket{u'v'}\bra{uv} + h.c. ) \}
\otimes\ket{a}. \nonumber \\
\end{eqnarray}
Once the reduced density matrix is so obtained and normalized, the entanglement can be quantified by von Neumann entropy, as usual.

\section{Entanglement of a two-particle basis state $\ket{i,j}$}
Here we calculate the entanglement of a basis state $\ket{i,j}$ within the complete single-particle basis, showing that it depends on the scalar product between $\ket{i}$, $\ket{j}$. We thus consider the two-identical particle state \mbox{$\ket{\Psi^{(2)}}=\ket{i,j}$}, where $\ket{i}$ and $\ket{j}$ are generic single-particle states. The reduced single-particle density matrix, performed on the basis $\{ \ket{i'} \}$, reads
\begin{equation*}
\rho^{(1)}=\frac{1}{2}\mathrm{Tr^{(1)}}\ket{\Psi^{(2)}}\bra{\Psi^{(2)}}=\frac{1}{2}\sum_{i'}\braket{i'|i,j}\braket{i,j|i'}.
\end{equation*}
By using Eq.~(2) of the manuscript, we obtain
\begin{align*}
\rho^{(1)}&
=\frac{1}{2}\{\bra{i}\sum_{i'} \ket{i'}\braket{i'|i}\ket{j}\bra{j}+\bra{j}\sum_{i'} \ket{i'}\braket{i'|j}\ket{i}\bra{i}+\\
&+\eta [\bra{j}\sum_{i'} \ket{i'}\braket{i'|i}\ket{j}\bra{i}+\bra{i}\sum_{i'} \ket{i'}\braket{i'|j}\ket{i}\bra{j}] \}.
\end{align*}
Recognizing the presence of the single-particle identity $\sum_{i'}\ket{i'}\bra{i'}=\mathbb{I}$, the reduced density matrix $\rho^{(1)}$ reduces to
\begin{equation}
\rho^{(1)}=\frac{1}{2} \{ \ket{j}\bra{j}+\ket{i}\bra{i}+\eta(\braket{j|i}\ket{j}\bra{i}+\braket{i|j}\ket{i}\bra{j}) \}.
\end{equation}
If $\ket{i}$ and $\ket{j}$ are orthogonal, i.e. $\braket{i|j}=0$, one obtains 
\begin{equation}
\rho^{(1)}=\frac{1}{2}(\ket{i}\bra{i}+\ket{j}\bra{j}).
\end{equation}
Since the eigenvalues of $\rho^{(1)}$ are $\lambda_1=\lambda_2=1/2$, entanglement as quantified by the von Neumann entropy is maximum (see Eq.~(6) of the main text). Moreover, the presence of entanglement (independently of its amount) is witnessed by the number of the nonvanishing eigenvalues, that is by the Schmidt number. 

On the other side, when the states $\ket{i}$, $\ket{j}$ coincide, that is $\braket{i|j}=1$, one obtains
\begin{equation}
\rho^{(1)}=\frac{1}{2}(1+\eta)\ket{i}\bra{i}.
\end{equation}
Such a condition, allowed only for bosons ($\eta=+1$), leads to $\rho^{(1)}=\ket{i}\bra{i}$ which is a pure state whose unique nonvanishing eigenvalue is $\lambda_1=1$. Entanglement of this state is thus zero (the von Neumann entropy vanishes), as already witnessed by the presence of only one nonvanishing eigenvalue in the reduced density matrix.

We then conclude that the entanglement of a two-particle basis state $\ket{i,j}$ depends on the scalar product (and thus on the angle, from a geometrical viewpoint) between the single-particle states $\ket{i}$ and $\ket{j}$. It is maximum when they are orthogonal and zero when they are the same. Furthermore, we have here confirmed that the Schmidt number is an entanglement witness, as for distinguishable particles \cite{preskill1998lecture}. We remark that these results are valid when the partial trace is performed within the complete single-particle basis in the specific case when the single-particle state is described by an observable alone. We have already seen (see section above and the case of two spatially separated particles of the manuscript) that for a single-particle state described by a number of observables, new scenarios surface for determining the entanglement of the identical particle system.

\section{Schmidt decomposition of the Bell-like state starting from a global partial trace}
We give here the Schmidt decomposition of the Bell-like state $\ket{\Psi}$ of Eq.~(7) of the manuscript, by following the recipe in the main text. We first perform the global partial trace of $\rho$ on the total single-particle space $\{ \ket{L\uparrow}, \ket{L\downarrow}, \ket{R\uparrow}, \ket{R\downarrow} \}$, and obtain
\begin{eqnarray}
\rho^{(1)}=\frac{1}{2}(\alpha^2\Ket{L\uparrow}\Bra{L\uparrow}+|\beta|^2\Ket{L\downarrow}\Bra{L\downarrow}+\nonumber\\
+|\beta|^2\Ket{R\uparrow}\Bra{R\uparrow}+\alpha^2\Ket{R\downarrow}\Bra{R\downarrow}).
\end{eqnarray}
It has eigenvalues $\lambda_1=\lambda_4=\alpha^2/2$, $\lambda_2=\lambda_3=|\beta|^2/2$, and eigenstates $\ket{1}=\eta\ket{\tilde{4}}=\Ket{L\uparrow}$, $\ket{\tilde{1}}=\Ket{4}=\ket{R\downarrow}$, $\Ket{2}=\eta\ket{\tilde{3}}=\ket{L\downarrow}$, $\ket{\tilde{2}}=\ket{3}=\ket{R\uparrow}$, which define the Schmidt basis. The Schmidt decomposition of the Bell-like state thus results
\begin{equation}
\ket{\Psi}=\frac{1}{\sqrt{2}}\left[\frac{\alpha}{\sqrt{2}}(\ket{1,\tilde{1}}+\ket{4,\tilde{4}})
+\frac{|\beta|}{\sqrt{2}}(\ket{2,\tilde{2}}+\ket{3,\tilde{3}})\right],
\end{equation}
and it permits to write the von Neumann entropy $S(\rho^{(1)})$ of Eq.~(11) of the manuscript. Notice the difference between the SD given here and that reported in Eq.~(9) of the main text.

\section{Schmidt decomposition of the two-boson state $\ket{\Phi}$}
We here provide the Schmidt decomposition of the two-boson state $\ket{\Phi}=\ket{\uparrow,\uparrow_\textbf{u}}$ defined in Eq.~(12) of the manuscript. The reduced density matrix $\rho^{(1)}$, obtained by performing the partial trace of $\rho=\ket{\Phi}\bra{\Phi}$ on the basis $\{\ket{\uparrow}, \ket{\downarrow} \}$, is
\begin{equation}
\rho^{(1)}=\frac{1}{2\mathcal{N}}
\begin{pmatrix}
a&c\\
c^\ast&b
\end{pmatrix},
\end{equation}
where $a=4\cos^2\frac{\theta}{2}+\sin^2\frac{\theta}{2}$, $b=\sin^2\frac{\theta}{2}$, $c=e^{i\phi}\sin\theta$ and $\mathcal{N}=(1+\cos^2\frac{\theta}{2})$. 
It is straightforward to find its eigenvalues
\begin{equation}
\lambda_1=\frac{4}{\mathcal{N}}\cos^4\frac{\theta}{4},\quad  \lambda_2=1-\lambda_1=\frac{4}{\mathcal{N}}\sin^4\frac{\theta}{4},
\end{equation}
and the corresponding eigenstates
\begin{equation}
\ket{1}=\cos\frac{\theta}{4}\ket{\uparrow}+\sin\frac{\theta}{4}\ket{\downarrow},\
\ket{2}=i(-\sin\frac{\theta}{4}\ket{\uparrow}+\cos\frac{\theta}{4}\ket{\downarrow}).
\end{equation}
As we see, they only depend on the angle between the pseudospins ($\theta$). Since we are dealing with two bosons in the same site, whose single-particle states are described by only an observable (the pseudospin), and the eigenvalues are nondegenerate, the single-particle states $\ket{i}$, $\ket{\tilde{i}}$ defining the Schmidt basis $\ket{i,\tilde{i}}$ are $\ket{1}=\ket{\tilde{1}}$ and $\ket{2}=\ket{\tilde{2}}$. Therefore, the (normalized) Schmidt decomposition of the state $\ket{\Phi}$, obtained by Eq.~(3) of the manuscript, is given by
\begin{equation}
\ket{\Phi}=\frac{1}{\sqrt{2}}(\sqrt{\lambda_1}\ket{1,\tilde{1}}+\sqrt{\lambda_2}\ket{2,\tilde{2}}). 
\end{equation}
The corresponding entanglement is quantified by the von Neumann entropy $S(\rho^{(1)})=-\sum_{i=1}^2\lambda_i\log_2\lambda_i$.

\section{Schmidt decomposition of the state $\ket{\Psi_\phi}$ of two qutrits in the same site}
We give the Schmidt decomposition of two identical qutrits in the same site, each characterized by the basis $ \{ \ket{e_1}, \ket{e_2}, \ket{e_3} \} $. This system is equivalent to that of two spin-1 bosons in the same hole, previously analyzed by an alternative method \cite{balachandran2013entanglement}. We consider the state
\begin{equation}\label{stato balach}
\ket{\Psi_\phi}=\cos\phi\ket{e_1,e_2}+\sin\phi\ket{e_1,e_3},
\end{equation}
where the spatial index has been omitted for simplicity. By performing the partial trace of $\rho$ onto the basis $ \{ \ket{e_1}, \ket{e_2}, \ket{e_3} \} $, we obtain the reduced density matrix
\begin{equation}
\rho^{(1)}=\frac{1}{2}
\left(\begin{matrix}
1&0&0\\
0&\cos^2\phi&\sin\phi\cos\phi\\
0&\sin\phi\cos\phi&\sin^2\phi
\end{matrix}\right),
\end{equation}
which has eigenvalues $\lambda_1=\lambda_2=1/2$, $\lambda_3=0$ and eigenstates $\ket{1}=\ket{\tilde{2}}=\cos\phi\ket{e_2}+\sin\phi\ket{e_3}$, $\ket{\tilde{1}}=\ket{2}=\ket{e_1}$, $ \ket{3}=-\sin\phi\ket{e_2}+\cos\phi\ket{e_3}$, $\ket{\tilde{3}}=0$, which define the Schmidt basis. From Eq.~(3) of the main text, the SD of the state is
 \begin{equation}\label{schmidt balach}
\ket{\Psi_\phi}=\frac{1}{\sqrt{2}}\left(\frac{1}{\sqrt{2}}\ket{1,\tilde{1}}+\frac{1}{\sqrt{2}}\ket{2,\tilde{2}}\right).
\end{equation}

\end{document}